# Propagation of Bessel and Airy beams through atmospheric turbulence

W. Nelson*, J.P. Palastro**, C.C. Davis*, and P. Sprangle*,**

*Department of Electrical and Computer Engineering, University of Maryland
College Park, Maryland20740
**Institute for Research in Electronics and Applied Physics, University of Maryland
College Park, Maryland20740

**Abstract**

We investigate, through simulation, the modifications to Bessel and Airy beams during propagation through atmospheric turbulence. We find that atmospheric turbulence disrupts the quasi-non-diffracting nature of Bessel and Airy beams when the transverse coherence length (Fried parameter) nears the initial aperture diameter or diagonal respectively. The turbulence induced transverse phase distortion limits the effectiveness of Bessel and Airy beams for applications requiring propagation over long distances in the turbulent atmosphere.

# I. Introduction

Phase distortions acquired from turbulent fluctuations in the refractive index modify the propagation of laser beams through atmosphere [1-7]. In weak atmospheric turbulence, the modifications can be described as three separate effects: wander, spreading, and scintillation each of which can be detrimental to applications requiring long range propagation of the beam [4,7]. For instance, beam wander, or deflections in the beam's centroid due primarily to the large scale fluctuations, can result in the beam missing a target entirely. Beam spreading, expansion of the beam beyond vacuum diffraction due primarily to small scale fluctuations, distributes the beam power over a larger area, reducing the intensity, delivered power, and efficiency. And scintillation, fluctuations in the beam's intensity, can result in image distortion or higher bit-error-rates in optical communication systems [8].

For many applications, the beams' initial transverse intensity profile has a single, on-axis peak such as a Gaussian, a 'flat-top' generated by a circular aperture, or some approximation thereof. Any additional peaks or non-monotonic decreases in the profile are often categorized as higher-order mode aberrations, and because they result in enhanced diffraction of the beam are usually undesirable [9]. Bessel and Airy beams are an exception. Bessel beams have cylindrically symmetric profiles with a central peak surrounded by concentric rings each possessing nearly the same amount of energy. During propagation the outer rings diffract inward fueling the center of the beam with energy and maintaining constant on-axis intensity. The inward diffraction also results in self-healing of the beam: when the beam's propagation path is partially obstructed the profile can nearly reform itself down-stream [10-14].

Airy beams, on the other hand, do not have cylindrical symmetry. The profile appears waffle-like with the intensity peaking in one corner and dropping with distance from the peak. While the Airy beam's centroid follows a straight trajectory, the peak intensity propagates along a curved path. These curved paths approximately follow the ballistic trajectory of projectiles and exhibit the same 'stalling' in motion that gravity produces in launched projectiles [14-17]. As with the Bessel beam, the interference of diffracting beamlets in the Airy beam's waffle-like pattern results in self-healing [14].

True Bessel and Airy beams are, however, only theoretical. Like plane waves, their field profiles would need to extend to infinity, requiring an unphysical, infinite amount of energy. In practice, the transverse profile of Bessel and Airy beams must be truncated at a finite radius or 'super-apertured,' for instance with a Gaussian profile, changing how the beam propagates [11]. Nevertheless, even when apertured, the desirable features of Bessel and Airy beams survive over extended distances in vacuum. As a result, it has been suggested that Bessel beams [18] and Airy beams [14,19] could be useful for remote sensing or directed energy applications. If the desirable properties of Bessel and Airy beams are to be harnessed for these applications, they must be robust to phase distortions resulting from atmospheric turbulence [20-23].

Here we investigate, through the use of propagation simulations, how turbulence modifies the propagation of Bessel and Airy beams. For both beams, we find the extent that each beam is modified by atmospheric turbulence depends on the transverse beam size. The transverse coherence length (Fried parameter), describing the transverse distance over which the phase fronts of the laser beam remain correlated, decreases with propagation distance. As a result, large diameter beams acquire transverse phase distortions that modify their propagation before they undergo standard diffraction. Small diameter beams, on the other hand, diffract before their transverse coherence is disrupted. Furthermore we find that the nature of turbulence-induced beam spreading differs between Bessel and Airy beams and Gaussian beams: the simultaneous diffraction of many rings or beamlets make the Bessel and Airy beams resistant to spreading in turbulence. Despite the Bessel beams resistance to spreading, the power delivered to the initial aperture area decreases as the number of rings increases. This suggests that the most effective Bessel beam for delivering power is a Bessel beam with zero rings, essentially a Gaussian beam.

In section II, we describe a model for propagation through turbulence. Section III discusses the simulation technique, namely a pseudo-spectral propagation algorithm for the paraxial equation with turbulence-induced refractive index fluctuations included through phase screens. Section IV presents results for propagation of Gaussian, Bessel, and Airy beams through turbulence and simple scalings to describe the observed phenomena. Section V ends the manuscript with a summary of our results and conclusions.

## II. Propagation through turbulence

We write the transverse electric field of the laser beam as a plane wave carrier modulating an envelope as follows:

$$E_\perp(\mathbf{r},t) = \frac{1}{2}\hat{E}_\perp(\mathbf{r},t)\exp[i(kz-\omega_0 t)] + c.c., \quad (1)$$

where $k = k_0 + \delta k$, $k_0 = 2\pi/\lambda_0$ and $\omega_0 = ck_0$ are the carrier wave number and central frequency of the laser beam respectively and $\delta k$ is to be determined. We limit our investigation to beams with duration long enough and power low enough such that the envelope is independent of time. The evolution of the beam envelope is then determined by the steady-state, enveloped paraxial equation

$$\left[\nabla_\perp^2 + 2ik_0\frac{\partial}{\partial z}\right]\hat{E}_\perp(\mathbf{r})e^{i\delta kz} = -2k_0^2\delta n\hat{E}_\perp(\mathbf{r})e^{i\delta kz} \quad (2)$$

where $\delta n = n(\omega_0) - 1$ is the shift in refractive index accounting for linear dispersion. In atmosphere the refractive index consists of a mean contribution and a fluctuating contribution due to turbulence, $\delta n = \langle \delta n \rangle + \delta n_T(\mathbf{r})$ with $\langle \delta n_T(\mathbf{r}) \rangle = 0$. The average, $\langle \ \rangle$, is performed over an ensemble of statistically independent instances of the index fluctuations. We choose $\delta k = \langle \delta n \rangle k_0$ to remove the mean index from Eq. (2), providing

$$\left[\nabla_\perp^2 + 2ik_0\frac{\partial}{\partial z}\right]\hat{E}_\perp(\mathbf{r}) = -2k_0^2\delta n_T(\mathbf{r})\hat{E}_\perp(\mathbf{r}). \quad (3)$$

The refractive index fluctuations, $\delta n_T(\mathbf{r})$, arise from temperature fluctuations. We note that strictly speaking the refractive index depends on the atmospheric density, but because the fluctuations are nearly isobaric, the density fluctuations are directly proportional to the temperature fluctuations. In the Kolmogorov cascade theory of turbulence, temperature fluctuations are formed with large scales sizes, defined as the outer scale $L_0$, corresponding to the scale over which the air is heated [3]. The fluctuations continually dissipate to smaller scales through molecular diffusion. The cascading to smaller scales occurs because the rate of dissipation increases as the fluctuation size decreases. When the dissipation rate equals the heating rate, the cascade terminates and defines the inner scale length of the fluctuations, $\ell_0$.

The refractive index fluctuations are characterized by their covariance function, $B_n(\mathbf{r},\mathbf{r}') = \langle \delta n_T(\mathbf{r}) \delta n_T(\mathbf{r}') \rangle$ and their structure function, $D_n(\mathbf{r},\mathbf{r}') = \langle [\delta n_T(\mathbf{r}) - \delta n_T(\mathbf{r}')]^2 \rangle$ [4]. We consider homogeneous, isotropic, Gaussian fluctuations such that $B_n(\mathbf{r},\mathbf{r}') = B_n(|\mathbf{r} - \mathbf{r}'|)$ or $D_n(\mathbf{r},\mathbf{r}') = 2[B_n(0) - B_n(|\mathbf{r} - \mathbf{r}'|)]$ fully determine the statistical properties. The Fourier transform of the covariance function, $\Phi_n(\kappa)$, represents the distribution of fluctuation scale sizes. Here we use the modified Von Karman spectrum

$$\Phi_n(\kappa) = 0.33 C_n^2 \frac{e^{-(\kappa \ell_0 / 2\pi)^2}}{(\kappa^2 + L_0^{-2})^{11/6}}, \quad (4)$$

where $C_n^2$ is the refractive index structure constant. Using Eq. (4), the index structure function can be shown to be $D_n \simeq C_n^2 |\mathbf{r} - \mathbf{r}'|^{2/3}$ for $\ell_0 \ll |\mathbf{r} - \mathbf{r}'| \ll L_0$ [3].

To assess the modifications to a laser beam in atmospheric turbulence, it is common to consider the solution to Eq. (3) in the Frauhofer diffraction limit. In particular, the ensemble averaged, on-axis intensity, $\langle I(0,z) \rangle = \frac{1}{2} c \varepsilon_0 \langle |\hat{E}_\perp(0,z)|^2 \rangle$, far from the aperture can be expressed

$$\langle I(0,z) \rangle = \frac{1}{2} c \varepsilon_0 \left( \frac{k_0}{2\pi z} \right)^2 \iint \hat{E}(\mathbf{r}'_\perp, 0) \hat{E}^*(\mathbf{r}''_\perp, 0) e^{-\frac{1}{2} \langle [\phi(\mathbf{r}'_\perp, z) - \phi(\mathbf{r}''_\perp, z)]^2 \rangle} d\mathbf{r}'_\perp d\mathbf{r}''_\perp \quad (5)$$

where

$$\phi(\mathbf{r}_\perp, z) = k_0 \int_0^z \delta n_T(\mathbf{r}_\perp, z') dz'.$$

The exponential in Eq. (5a) includes the phase structure function $D_s(|\mathbf{r}'_\perp - \mathbf{r}''_\perp|) = \langle [\phi(\mathbf{r}'_\perp, z) - \phi(\mathbf{r}''_\perp, z)]^2 \rangle$. Using the refractive index structure function, one can show that $D_s \simeq 2.91 k_0^2 C_n^2 z |\mathbf{r}_\perp - \mathbf{r}'_\perp|^{5/3}$ for $\ell_0 \ll |\mathbf{r} - \mathbf{r}'| \ll L_0$ [3,4]. With Eq. (5) and $D_s$, we can derive an estimate of the distance, $L_T$, at which turbulence becomes important. The ensemble averaged intensity drops by one e-folding due to turbulence when $D_s = 2$. Setting $|\mathbf{r}_\perp - \mathbf{r}'_\perp| = L_\perp$, the characteristic transverse variation in the

envelope, we find $L_T = 0.69(C_n^2 k_0^2 L_\perp^{5/3})^{-1}$. Up to a numerical factor this is the same distance at which the Fried parameter $r_0(z) = 1.67(C_n^2 k_0^2 z)^{-3/5}$ equals $L_\perp$ [3,4,7]. The Fried parameter, also referred to as the transverse coherence length, describes the transverse distance over which the phase fronts of the laser beam remain correlated. In other words, transverse distortions on the size of $r_0$ and modifications to beam propagation can be expected when the Fried parameter is smaller than the beam diameter.

## III. Simulation Description

### A. Simulation algorithm

For intervals of propagation, $\Delta z_s$, much shorter than the diffraction length, $L_d = \pi L_\perp^2 / \lambda$ ($L_\perp^2$ being the beam width of a Gaussian beam for example), the index fluctuations result in the accumulation of a transverse phase: $\hat{E}_\perp(\mathbf{r}_\perp, z + \Delta z_s) \simeq \hat{E}_\perp(\mathbf{r}_\perp, z) e^{i\phi(\mathbf{r}_\perp, \Delta z_s)}$, where

$$\phi(\mathbf{r}_\perp, \Delta z_s) = k_0 \int_z^{z+\Delta z_s} \delta n_T(\mathbf{r}_\perp, z') dz'. \quad (6)$$

If the propagation interval is also much larger than the outer scale, $\Delta z_s \gg L_0$, the accumulated phase can be approximated as [2,5]

$$\phi(\mathbf{r}_\perp, \Delta z_s) = (\pi \Delta z_s)^{1/2} k_0 \int_{-\infty}^{\infty} d\mathbf{\kappa}_\perp e^{i\mathbf{\kappa}_\perp \cdot \mathbf{r}_\perp} [a_r(\mathbf{\kappa}_\perp) + i a_i(\mathbf{\kappa}_\perp)] \Phi_n^{1/2}(\mathbf{\kappa}_\perp, 0), \quad (7)$$

where $a_r$ and $a_i$ are independent Gaussian random variables. Equation (7) provides the basis for the phase screen approximation in which the laser beam acquires the accumulated transverse phase at discrete axial points along the propagation path, and is propagated in vacuum between these points.

To simulate the propagation, we use a pseudo-spectral, split step algorithm. A typical axial advance, propagating the beam forward by $\Delta z$, involves three steps. In the first step, the transverse Fourier transform of the envelope is advanced a half step, $\Delta z / 2$ by applying the diffraction propagator: $\bar{E}_\perp(\mathbf{k}_\perp, z + \tfrac{1}{2}\Delta z) \simeq \bar{E}_\perp(\mathbf{k}_\perp, z) e^{i\frac{1}{2} k_0^{-1}(k_x^2 + k_y^2)\Delta z}$, where the overbar denotes the transform with respect to the transverse plane. For the second step, the phase screen is applied in coordinate space $E_\perp(\mathbf{r}_\perp, z + \tfrac{1}{2}\Delta z) \to E_\perp(\mathbf{r}_\perp, z + \tfrac{1}{2}\Delta z)e^{i\phi(\mathbf{r}_\perp,\Delta z)}$. Finally the beam is advanced the second half step in the transverse Fourier domain, $\bar{E}_\perp(\mathbf{k}_\perp, z + \Delta z) \simeq \bar{E}_\perp(\mathbf{k}_\perp, z + \tfrac{1}{2}\Delta z)e^{i\tfrac{1}{2}k_0^{-1}(k_x^2 + k_y^2)\Delta z}$. We note that a phase screen does not need to be applied at every advance but the interval between applications should satisfy the constraint that $L_0 \ll \Delta z_s \ll L_d$.

### B. Simulation details

We simulate the propagation of initially collimated laser beams with wavelengths of $\lambda_0 = 1\ \mu m$ through 6.4 $km$ of weak turbulence characterized by $C_n^2 = 1 \times 10^{-15}\ m^{-2/3}$, $\ell_0 = 1\ mm$, and $L_0 = 1\ m$. Phase screens are applied every 70 $m$ with the first screen applied at $z = 35\ m$. The transverse simulation domain is 1.2 $m$ by 1.2 $m$ with $N_\perp = 2048$ cells in each direction unless otherwise stated. The transverse scale length was varied and is discussed specifically for each beam profile in the next section.

The maximum propagation distance was chosen equal to the distance at which the Rytov variance, $\sigma_R^2(z) = 10.5 C_n^2 \lambda^{-7/6} z^{11/6}$, is unity. Longer propagation distances are considered the regime of 'strong' optical turbulence where the phase screen method may no longer apply [3,4]. Specifically the Rytov variance quantifies the intensity variance, $\sigma_I^2 = \langle I^2 \rangle / \langle I \rangle^2 - 1$, of a plane wave propagating through turbulent atmosphere. For weak optical turbulence, $\sigma_R^2 < 0.5$, $\sigma_I^2 \simeq \sigma_R^2$. To validate the simulation, the linear scaling of intensity variance with Rytov variance was confirmed up to $\sigma_R^2 \sim 0.5$ for a 100 $cm$ width Gaussian beam in a simulation domain 4 $m$ on a side with $N_\perp = 2048$. The intensity variance was calculated in a circle of radius 30 $cm$ centered at the beam centroid. The choice of beam width and circle radius ensured that the intensity variance was calculated in a region were the beam propagation resembled plane wave propagation.

## IV. Results

## A. Gaussian Beam

We begin by reviewing the propagation of Gaussian beams through vacuum and turbulence. The Rayleigh range, $L_R = \pi w_0^2 / \lambda$ where $w_0$ is the initial $1/e$ field width, defines the length scale over which a Gaussian beam diffracts in vacuum. In particular, the on-axis intensity of a Gaussian beam drops, $I(0,z) = I_0[1+(z/L_R)^2]^{-1}$ with a concurrent increase in the spot size, $w(z) = w_0[1+(z/L_R)^2]^{1/2}$. As discussed above, turbulence has a significant effect on beam propagation when the transverse coherence length approaches the beam diameter. Setting $r_0 = 2w_0$, we find an approximate length scale over which turbulent spreading and wander will cause a drop in the on-axis intensity: $L_T \sim 0.74(C_n^2 k_0^2 w_0^{5/3})^{-1}$. By comparing $L_R$ and $L_T$, the relative importance of diffractive spreading and turbulence induced modifications to the beam can be ascertained. For example, if $L_R \gg L_T$ turbulence will strongly modify the beam before significant diffraction occurs.

For a qualitative view of how atmospheric turbulence modifies the beam we turn to Fig. 1. The figure displays the intensity profile of a Gaussian beam with initial spot size $w_0 = 4.5$ $cm$ after 6.4 km of propagation in (a) vacuum, (b) turbulence, and (c) averaged over 100 instances through turbulence. The color scales are normalized to the maximum in each plot. A near 5 cm wander of the beam centroid and spreading can be observed in Fig 1(b). Figure 1(c) shows that, when ensemble averaged, the beam's width increases more rapidly than in vacuum. These modifications are expected as $L_R = 6.4$ $km$ is larger than $L_T = 3.3$ $km$.

A more quantitative examination and scaling with $w_0$ is presented in Fig. (2). Results from propagation through turbulence are represented by red, from propagation through vacuum by blue, and the ratio of Fried parameter to beam diameter, $r_0/2w_0$, is displayed in green. Figures 2(a) and 2(b) display 100 run ensemble averages of the normalized on-axis intensity, $I(0,z)/I_0$, as a function of propagation distance for initial spot sizes of 1.8 $cm$ and 4.5 $cm$ respectively. The maximum of 4.5 $cm$ was chosen such that the Rytov variance reached unity at one Rayleigh range. For $w_0 = 1.8$ $cm$, turbulence has minimal effect on the beam's on-axis intensity as the ratio of Fried parameter to beam diameter remains greater than unity over several Rayleigh ranges, $L_T = 15 \ km$ while $L_R = 1.0 \ km$. In contrast, when $w_0 = 4.5 \ cm$, $r_0/2w_0 = 1$ after $3.3 \ km$ well before the beam has propagated an entire Rayleigh range, $L_R = 6.4 \ km$. As a result, the on-axis intensity is ~20% lower than the vacuum value after 6.4 km of propagation.

By forming the ratio $L_T / L_R \propto w_0^{-11/3}$, we see that smaller beams are susceptible to intensity loss through standard vacuum diffraction, while larger beams are susceptible to spreading and wander in turbulence. In particular, noticeable turbulent modifications should occur when $L_T \sim L_R$ or $w_0 \sim 3.8 \ cm$. This trend is illustrated in Figs. 1(c) and 1(d). Figure 1(c) displays the fractional increase in root mean square (RMS) radius, $w_{rms}/w_0$, after one Rayleigh length as a function of $w_0$. For reference the initial value is $w_{rms}(0)/w_0 = 1/\sqrt{2}$. In vacuum, $w_{rms}/w_0$ increases by a factor of $\sqrt{2}$ regardless of initial beam width as borne out by the blue curve. Nearly the same $\sqrt{2}$ increase is observed for smaller beams, $w_0 \sim 2 \ cm$, after propagation through atmospheric turbulence. But as the initial spot size increases, the effect of turbulence on the beam size becomes sizeable: the $w_0 \sim 4.5 \ cm$ beam expands to ~1.5 its vacuum width. Figure 1(d) shows the on-axis intensity after one Rayleigh length of propagation normalized to its initial value, $I(0, L_R)/I_0$, as a function of $w_0$. Corresponding to the increase in RMS radius, the on-axis intensity has dropped by a factor of ~2 for all $w_0$ in vacuum and smaller $w_0$ in turbulence. As expected, the intensity of the larger beams has dropped significantly. However, the product of on-axis intensity and RMS radius squared has increased: $I(0, L_R)w^2(L_R) > I(0,0)w^2(0)$, suggesting that the beam has spread more from the edges than the center. Higher moments of the intensity distribution could be used to examine this effect, but we save this for future investigations.

**B. Bessel Beam**

Our goal now is to explore modifications to Bessel beam propagation in atmospheric turbulence. During vacuum propagation, each ring of the Bessel beam can be considered a separate beam undergoing its own diffraction. Each ring possesses near equal energy which it transports, through diffraction, outward from its initial radius and inward towards the center of the beam. The inward diffraction of the rings supplies the center of the beam with energy. This process maintains the on-axis intensity until the inward diffraction of the outer ring reaches the beam center. We can estimate the diffraction length of the Bessel beam by finding the distance at which the spot size associated with outer-ring expands to the total beam size. Noting that the Bessel function zeroes approach even spacing, we write the outer ring's spot size as $w_r \sim R/(N_r + 1)$ where $N_r$ is the number of rings and $R$ is the aperture radius. Setting $R = w_r[1 + (L_B/L_R)^2]^{1/2}$, where $L_R = \pi w_r^2/\lambda$, we find $L_B \sim \pi R^2/(N_r + 1)\lambda$. The diffraction length increases with aperture area and decreases with the number of rings.

A more precise length scale over which the Bessel beam diffracts is given by $L_B \sim 0.6\pi R^2/(N_r + 1)\lambda$, where the on-axis intensity of the Bessel beam will reach half its initial value, $I(0, L_B) \sim I_0/2$, after propagating a distance $L_B$. In the presence of turbulence, the rings lose spatial coherence and smear together, shortening the distance over which the on axis intensity remains constant. The length scale over which turbulence will cause a drop in the on-axis intensity can be approximated as: $L_T \sim 0.74(C_n^2 k_0^2 R^{5/3})^{-1}$. Similar to the Gaussian beam, if $L_B \gg L_T$ turbulence will modify the beam before significant diffraction occurs.

For Gaussian beams we found that the cause of on-axis intensity decay depended on the beam width. Roughly speaking, small and large beams were susceptible to standard diffraction and turbulence induced modifications respectively. Based on the ratio $L_T/L_B \propto (N_r + 1)^{-1} R^{-11/3}$, we expect Bessel beams to follow the same trend for fixed $N_r$. To examine this, we fix the number of rings at 14 and vary the initial aperture radius from 6.65 cm to 23 cm. The beams are hard-apertured between adjacent intensity rings at the zeros of the Bessel function. The maximum aperture radius was chosen such that the Rytov variance was approximately unity at a distance of $L_B$.

Figure 3 displays intensity profiles of the largest aperture, $R = 23 \ cm$, beam after 6.4 km of propagation in (a) vacuum, (b) turbulence, and (c) averaged over 100 instances through turbulence. The color scales are normalized to the maximum in each plot. For these parameters, $L_B = 6.5\ km$ is much larger than $L_T = 0.22\ km$. Figure 1(b) shows that the concentric, closed ring structure observed in Fig. 1(a) has been significantly distorted by atmospheric turbulence. The rings are no longer spatially coherent with themselves or the entire beam. After 6.4 km of propagation, the transverse coherence length is $r_0 \simeq 1.6\ cm$ similar to the observed scale length of intensity fluctuations.

In Figure 4 the results are demarcated as before: propagation through turbulence is represented by red, propagation through vacuum by blue, and the ratio of Fried parameter to beam diameter, $r_0/2R$, is displayed in green. Figures 4(a) and 4(b) display 100 run ensemble averages of the normalized on-axis intensity as a function of propagation distance for initial aperture radii of 6.65 $cm$ and 23 $cm$ respectively. The ripples in intensity result from the initial hard-aperturing. The $R = 6.65\ cm$ beam undergoes almost no loss of on-axis intensity due to turbulence before diffracting: $L_B = 0.55\ km$ is smaller than $L_T = 1.7\ km$. With an aperture radius of $R = 23\ cm$, the diffraction length far exceeds the turbulent modification length, $L_B \gg L_T$, and the beam's on-axis intensity drops to 90% of its vacuum value after 6.4 km of propagation.

Comparisons of intermediate apertures radii are provided in Figs. 4(c) and (d). Figure 4(c) displays the RMS radius normalized to the aperture radius at a distance $L_B$ as a function of aperture radius. For reference the value at the aperture is $w_{rms}(0)/R = 0.57$. Unlike the Gaussian beam, the Bessel beam's normalized RMS radius is equal for all initial apertures after propagating a diffraction length through turbulence. The rings undergo turbulent wander and spreading, which transports energy both away from and towards the center of the beam and fills the intensity gaps between the rings. Again we treat each ring as an individual beam with spot size $w_r \sim R/(N_r + 1)$. The total beam will undergo significant spreading in turbulence after the individual rings do so: distances $L > L_{T,r} \sim 0.74(C_n^2 k_0^2 w_r^{5/3})^{-1}$ with $N_r \gtrsim 3$ and where the subscript $r$ refers to the individual ring. For apertures of $R = 6.65\ cm$ and $R = 23\ cm$, $L_{T,r} = 160\ km$ and $L_{T,r} = 20\ km$ respectively. While the redistribution of energy within the beam does not affect its overall spreading, it does result in the decay of on-axis intensity as displayed in Fig. 4(d). As expected, the intensity decay is most severe for the larger beams, and turbulent modifications become apparent when $L_T \sim L_B$ or $R = 9\ cm$. Thus, as with Gaussian beams, small and large beams are susceptible to standard diffraction and turbulence induced spreading and wander respectively.

**C. Fixed Aperture Comparison**

In the previous section the effect of beam size on atmospheric propagation of fixed, 14 ring Bessel beams was examined. We now fix the beam size and power, and vary the number of rings. We chose an aperture size of $30\ cm$, approximating the size of a beam director for directed energy applications [7]. As before, the beams are hard-apertured between adjacent intensity rings at the zeros of the Bessel function. The top plot of Fig. (5) shows the fractional power delivered to a $30\ cm$ aperture at distances of $1.6\ km$ in blue, $4.0\ km$ in red, and $6.4\ km$ in green as a function of rings in the beam. The results in vacuum are represented by the dashed curve and turbulence by the solid curve. The delivered powers in vacuum and turbulence are nearly identical. As discussed above, over these distances, turbulence results in energy spreading within the beam without increasing the overall RMS radius beyond diffractive spreading. In particular, the distances over which turbulence modifies the beam spreading for $N_r = 3$ and $N_r = 14$ are $L_{T,r} = 4.5\ km$ and $L_{T,r} = 40\ km$ respectively. For all three distances, the delivered power decreases as the number of rings increases. This implies, for the situation considered here, that the most effective beam for power delivery is a Bessel beam with one zero, essentially a Gaussian beam clipped at the radius of the beam director.

The bottom plot of Fig. (5) shows the normalized on-axis intensity as a function of rings in the beam. The distances are the same as above. With fixed power and aperture area, the initial on-axis intensity is given by $I_0 \simeq (\pi/2)(N_r + 1.2)R^{-2}P$. Thus before significant diffraction or turbulent spreading and wander, the on-axis intensity increases with number of rings as demonstrated by the blue-dashed curve. The ratio $L_T / L_B \propto (N_r + 1)^{-1}$ indicates that a beam's sensitivity to on-axis intensity loss in turbulence increases with its number of rings. As illustrated by the red and green curves, the decay in on-axis intensity relative to the vacuum values increases with $N_r$. The result of these two effects is that the ideal beam profile for applications requiring high peak intensity depends on the distance to the target.

**D. Airy Beam**

We now explore modifications to Airy beam propagation in atmospheric turbulence. Like the rings of the Bessel beam, each beamlet of the Airy beam can be considered a separate beam undergoing its own diffraction. During propagation through vacuum, the beamlets interfere such that the Airy beam's center of mass follows a straight line while its intensity maximum follows a parabolic trajectory, drifting in the transverse plane. In particular, the transverse position of the intensity peak, $\mathbf{r}_{\perp,p}$, evolves according to $\mathbf{r}_{\perp,p} = \mathbf{r}_{\perp,i} + k_0^{-2} z^2 / w_A^3 \hat{\mathbf{r}}_\perp$ where $\mathbf{r}_{\perp,i}$ is the initial position of the peak and $w_A$ is the Airy function scaling constant, ie. $Ai(x/w_A)Ai(y/w_A)$. We can estimate the Airy beam's diffraction length, $L_A$, by finding the distance at which an individual beamlet expands to the total beam size. To proceed, we define $N_{bl}^2$ as the total number of beamlets, and $D$ as the distance from the origin to the outer edge of the $N_{bl}^{th}$ beamlet along a single cartesian direction. Said differently, the function $Ai(x/w_A)$ has $N_{bl}$ zeros with the last zero at $x = D$. Using a typical beamlet of width $w_{bl} \sim D/N_{bl}$, and following the same arguments used for the Bessel beam above, we find $L_A \sim D^2 / N_{bl} \lambda$. The diffraction length increases with the transverse beam area and decreases with the number of beamlets.

The more precise length scale over which the airy beam diffracts is given by $L_A \sim 2D^2 / N_{bl} \lambda$. In particular the peak intensity of the Airy beam reaches half its initial value at a distance $L_A$: $I(x_p, y_p, L_A) \sim I(x_p, y_p, 0)/2$. As with Gaussian and Bessel beams, we can compute the distance at which turbulence affects the Airy beam by equating the Fried parameter with the largest transverse dimension of the beam.

Approximating the Airy beam as a square, the largest dimension is the diagonal $d = 2^{1/2} D$. Setting $r_0 = d$ we find $L_T \sim 1.2 (C_n^2 k_0^2 D^{5/3})^{-1}$.

We expect the Airy beams to exhibit the same decay in maximum intensity as the Gaussian and Bessel beams experienced in on-axis intensity. Furthermore the ratio $L_T / L_A \propto N_{bl}^{-1} D^{-11/3}$ has a similar scaling as that of the Bessel beam, suggesting that small and large Airy beams should be susceptible to standard diffraction and turbulent spreading and wander respectively. To examine the extent that turbulence modifies different size Airy beams, we fix $N_{bl}$ at 15 and vary $d$ from 11.3 cm to 31.1 cm.

Figure 6 displays intensity profiles for the largest beam, $d = 31.1$ cm, after 6.4 km of propagation in (a) vacuum, (b) turbulence, and (c) averaged over 100 instances through turbulence. The color scales are normalized to the maximum in each plot. At $z = 0$, the maximum intensity occurs at $x = y = -.012$ m. Figure 1(a) shows that the intensity peak has drifted diagonally to $x = y = .12$ m, while the tails of the beam have extended in the opposite direction, keeping the center of mass at a fixed transverse position. The distortion of the intensity profile due to turbulence is demonstrated in Fig. 1(b). The intensity peak occurs near the same location as in vacuum, but the individual beamlets have undergone wander, and some are no longer discernable having smeared together. For these parameters, the Fried parameter, $r_0 = 6$ cm, is still larger than a typical beamlet width, $w_{bl} \sim 1.5$ cm, while $L_A = 6.4$ km is much larger than $L_T = 0.38$ km.

Figure 7 uses the same color scheme as Figs. 2 and 4: propagation through turbulence is represented by red, propagation through vacuum by blue, and the ratio of Fried parameter to beam diagonal length, $r_0 / d$, is displayed in green. Figures 7(a) and 7(b) display 100 run ensemble averages of the normalized on-axis intensity as a function of propagation distance for the smallest and largest beam aperture diagonals, 11.3 cm and 31.1 cm respectively. As with the Bessel Beam, the ripples in maximum intensity are due to hard-aperturing. The smallest beam undergoes standard diffraction before being affected by turbulence: $L_A = 0.85$ km and $L_T = 2$ km. In contrast, turbulence has a pronounced effect on the on-axis intensity of the largest beam: $L_A = 6.4\ km$ and $L_T = 0.38\ km$.

Comparisons of intermediate aperture diagonal lengths are provided in Figs. 7(c) and (d). Figure 7(c) displays the ratio of RMS radius after one diffraction length to the initial aperture diagonal, illustrating the degree to which the Airy beams spreads. For reference, the initial value is $w_{rms}(0)/d = 0.44$. As with the Bessel beam, the Airy beam's normalized RMS radius is equal for all initial apertures after propagating a diffraction length through turbulence. The individual beamlets undergo turbulent wander and spreading filling the intensity gaps between them without increasing the overall beam size. Following the same argument presented above for Bessel beams, the total beam will undergo significant spreading in turbulence after the individual beamlets do so: distances $L > L_{T,bl} \sim 0.74(C_n^2 k_0^2 w_{bl}^{5/3})^{-1}$. For diagonals of $d = 11.3\ cm$ and $d = 31.1\ cm$, $L_{T,bl} = 115\ km$ and $L_{T,bl} = 21\ km$ respectively. Figure 7(d) shows that the maximum intensity of the Airy beams is affected by turbulence in the same way the on-axis intensity of Gaussian and Bessel beams are affected.

## V. Summary and Conclusions

We have investigated the propagation of Gaussian, Bessel, and Airy beams through atmospheric turbulence. The beam propagation was simulated using the paraxial wave equation with turbulence-induced refractive index fluctuations included through phase screens. The extent that each beam was modified by atmospheric turbulence depended on the transverse beam size. In particular, the transverse coherence length decreases with propagation distance: large aperture beams acquire transverse phase distortions before undergoing significant diffraction; small aperture beams diffract before acquiring significant transverse phase distortions. This trend held for all three beams and manifested in a drop in peak intensity. However, the nature of turbulence-induced beam spreading differed between the Bessel and Airy beams and the Gaussian beam. The simultaneous diffraction of many rings or beamlets made the Bessel and Airy beams resistant to spreading in turbulence. The propagation distance at which the Fried parameter becomes comparable to the ring or beamlet size far surpasses the distance at which it becomes comparable to the total transverse beam size.

We have also examined the scaling of power and intensity delivered to a target with the number of rings in a fixed aperture area Bessel beam. Despite the Bessel beams resistance to spreading, the power delivered to an area equal to the beam director decreases as the number of rings increases. As a result, the most effective Bessel beam for delivering power is a Bessel beam with zero rings, which approximates a Gaussian beam. Finally, the optimal number of rings for delivering on-axis intensity to a target depends on the distance to the target.

## Acknowledgements

The authors would like to thank T. Antonsen and H. Milchberg for fruitful discussions. This work was supported by funding from JTO, through ONR under Contract No. N000141211029, and AFOSR.

## References

[1] R.L. Fante, Proc. IEEE **63**, 1669 (1975).
[2] J.A. Fleck et al., Appl. Phys. **10**, 129 (1976).
[3] L.C. Andrews and R.L. Phillips, *Laser Beam Propagation through Random Media* (SPIE Press, Bellingham, WA, 2005).
[4] P.W. Milonni and J.H. Eberly, *Laser Physics* (John Wiley & Sons, Inc., Hoboken, NJ, 2010).
[5] R. Frehlich, Appl. Opt. **39**, 393 (2000).
[6] S.L. Chin et al., Appl. Phys. B **74**, 67 (2002).
[7] P.A. Sprangle et al., IEEE J. Quant. Electron. **45**, 138 (2009).
[8] C. C. Davis and I. I. Smolyaninov, Proc. SPIE 4489, 126 (2002).
[9] A. E. Siegman, OSA TOPS **17,** 185 (1998).
[10] J. Durnin, JOSA A **4**, 651 (1986).
[11] P. Sprangle and B. Hafizi, Phys. Rev. Lett. **66**, 837 (1991).
[12] J. Arlt and K. Dholakia, Optics Comm., Volume **177**, 297 (2000).
[13] L. Gong et al., Appl. Opt. **52,** 19 (2013).
[14] M. Bandres et al., Optics and Photonics **24**, 30 (2013).
[15] G.A. Siviloglou et al., Phys. Rev. Lett. **99**, 213901 (2007).
[16] G.A. Siviloglou et al., Opt. Letters **32**, 979 (2007).
[17] X. Ji et al., Opt.Express **21**, 2154 (2013).
[18] P. Polynkin et al.,Optics Express **16**, 15733 (2008).
[19] P. Polynkin et al., Science 324, 229 (2009).

[20] C. Bao-Suan and P. Ji-Xiong, Chinese Phys. B **18**, 1033 (2009).
[21] I. P. Lukin, Atmospheric and Oceanic Optics **25**, 328 (2012).
[22] Y. Gu and G. Gbur, Opt. Letters **35**, 3456 (2010).
[23] X. Chu, Opt. Letters **36**, 2701 (2011).

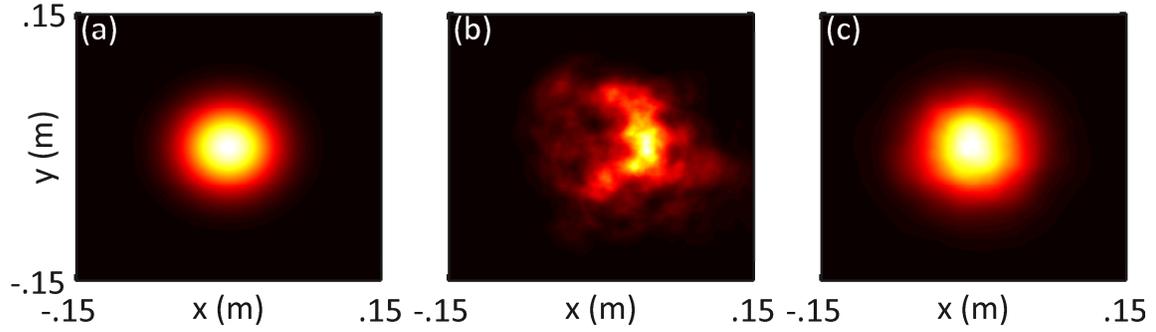

Figure 1. Transverse profiles of a Gaussian beam with initial spot size $w_0 = 4.5$ $cm$ after 6.4 km of propagation in (a) vacuum, (b) turbulence with $C_n^2 = 1 \times 10^{-15}$ $m^{-2/3}$, and (c) averaged over 100 instances through turbulence.

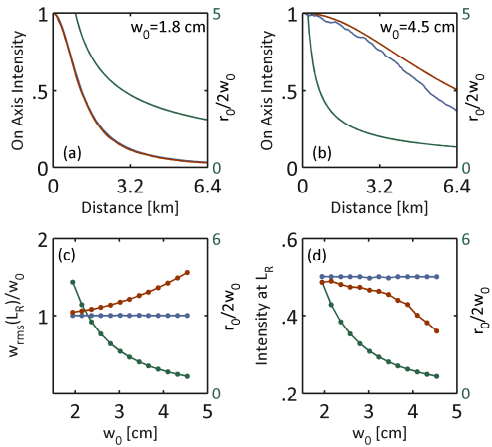

Figure 2. Simulation results for a Gaussian beam. (a) and (b) Comparison of the on-axis intensity in turbulence, red, and vacuum, blue, as a function of propagation distance for initial spot sizes of 1.8 cm and 4.5 cm respectively. (c) Ratio of the RMS radius at one Rayleigh length to the initial RMS radius as a function of initial spot size. (d) Normalized on axis intensity at one Rayleigh length as a function of initial spot size. In (a-d) the ratio of the Fried parameter to beam diameter, green, is plotted for reference.

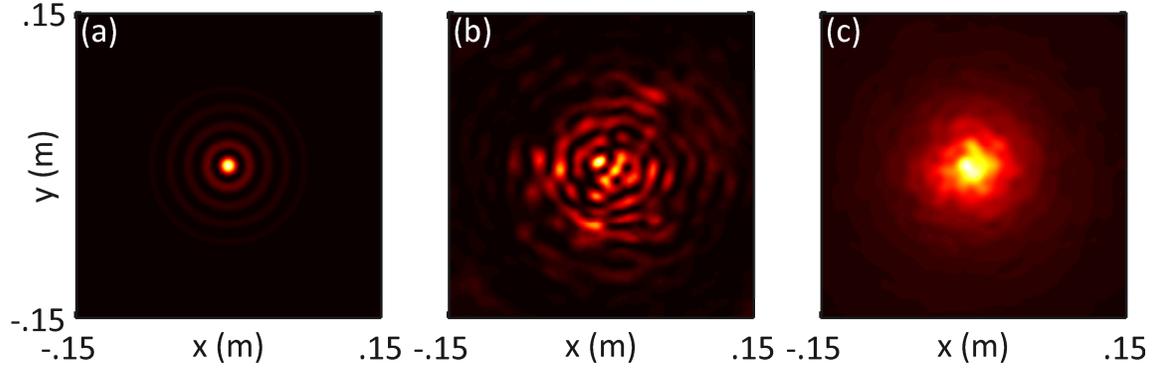

Figure 3. Transverse profiles of a 15 ring Bessel beam hard-apertured at a radius of 23 cm after 6.4 km of propagation in (a) vacuum, (b) turbulence with $C_n^2 = 1 \times 10^{-15}\ m^{-2/3}$, and (c) averaged over 100 instances through turbulence.

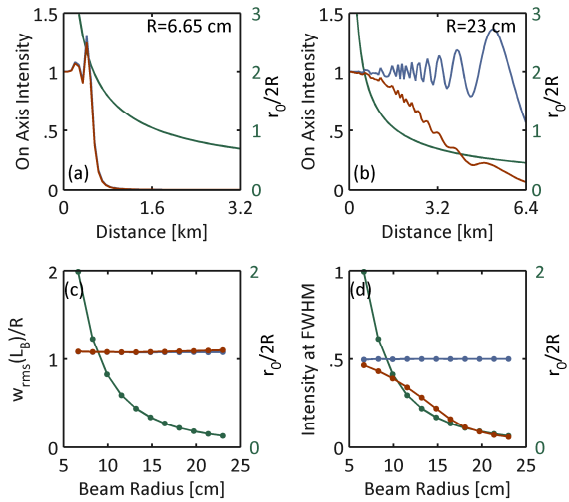

Figure 4. Simulation results for a Bessel beam. (a) and (b) Comparison of the on-axis intensity in turbulence, red, and vacuum, blue, as a function of propagation distance for initial aperture radii of 6.65 cm and 23 cm respectively. (c) Ratio of the RMS radius at one diffraction length, $L_B$, to the initial aperture radius as a function of initial aperture radius. (d) Normalized on axis intensity at one diffraction as a function of initial aperture radius. In (a-d) the ratio of the Fried parameter to beam diameter, green, is plotted for reference.

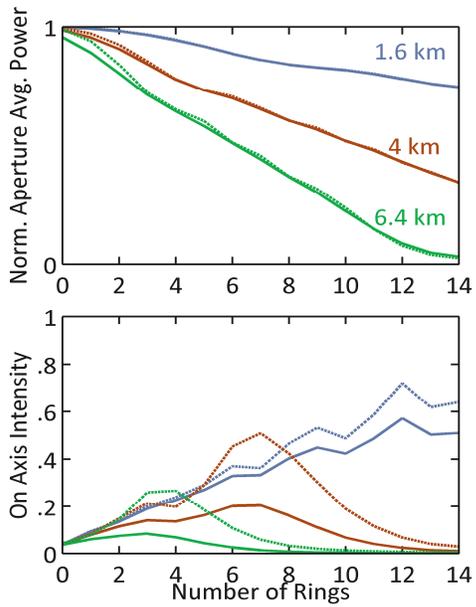

Figure 5. The power delivered to a circular aperture of radius 15 cm (top) and on axis intensity (bottom) as a function of rings in a Bessel beam initially apertured at a radius of 15 cm after propagation distances of 1.6 km, blue, 4 km, red, and 6.4 km, green. The dashed and solid lines are results from propagation in vacuum and turbulence with $C_n^2 = 1 \times 10^{-15} \ m^{-2/3}$ respectively.

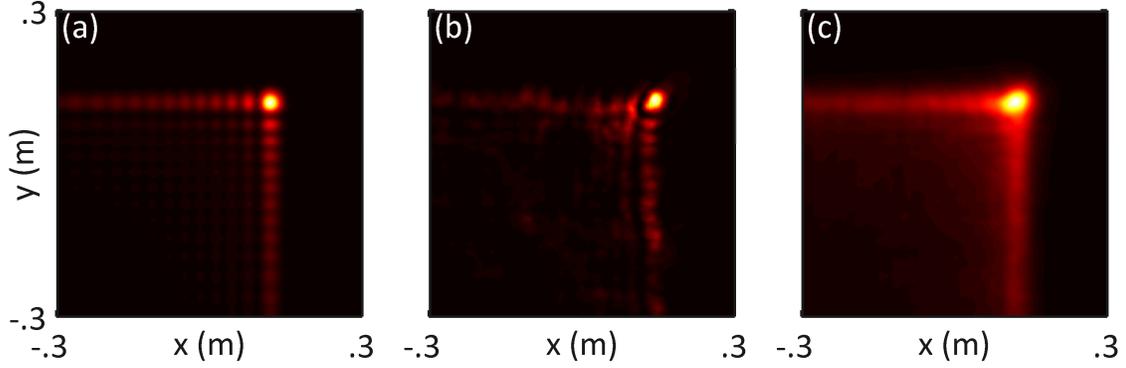

Figure 6. Transverse profiles of a 15 zero Airy beam hard-apertured at $y = -22$ $cm$ and $x = -22$ $cm$ after 6.4 km of propagation in (a) vacuum, (b) turbulence with $C_n^2 = 1 \times 10^{-15}$ $m^{-2/3}$, and (c) averaged over 100 instances through turbulence.

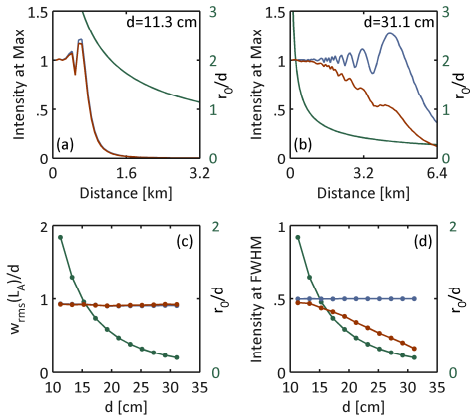

Figure 7. Simulation results for an Airy beam. (a) and (b) Comparison of the on-axis intensity in turbulence, red, and vacuum, blue, as a function of propagation distance for an aperture diagonal of 11.3 cm and 31.1 cm respectively. (c) Ratio of the RMS radius at one diffraction length, $L_A$, to the aperture diagonal as a function of aperture diagonal. (d) Normalized on axis intensity at one diffraction as a function of aperture diagonal. In (a-d) the ratio of the Fried parameter to beam diagonal, green, is plotted for reference.